# *A priori* Determination of the Extensional Viscosity of Polymer Melts

Nicholas A. Rorrer and John R. Dorgan*

*Author to whom correspondence should be addressed. jdorgan@mines.edu

KEYWORDS: Polymer Rheology, Extensional Flow, Dynamic Monte Carlo

ABSTRACT: The COMOFLO dynamic Monte Carlo algorithm is extended to enable the simulation of steady-state extensional rheology of dense melts for the first time ever. This significant advancement provides the ability to accurately capture the extensional viscosity of entangled polymer melts from low deformation rates, throughout the strain hardening regime, and into the region of viscosity thinning  To do so, the purely extensional flow field associated with a four roll mill experiment is implemented in three dimensions. Periodic boundary conditions exist in all directions to enable a uniform deformation rate throughout the simulation domain. As expected, the extensional viscosity is four times the zero shear viscosity at low deformation rates. As the deformation rate increases, the extensional viscosity increases, plateaus, and then decreases. The algorithm thus correctly captures the physics of polymer melts in extensional flow in a fully *a priori* manner while providing significant computational advantages over molecular dynamics methods. Because the technique is applicable to polydisperse systems, the ability to predict in an *a priori* manner both shear and elongational rheology for linear polymer melts of arbitrary molecular weight distribution can now be considered a solved problem.

The extensional behavior of polymer melts is of critical technological importance. [2, 3] Extensional rheology plays an important role in the flow of polymer solutions in microfluidics,[4, 5] in film blowing[6-8] and fiber spinning. [9] Given its importance, there are now a wide variety of instruments available to measure extensional properties (each with their own advantages and disadvantages) so that a large body of extensional data is available. [10-14]

Molecular simulations provide an attractive solution to understanding the experimentally observed behavior of a wide variety of polymeric materials. [15, 16] Recent advances allow for the simulation of polymer melts under flow, for example, the slip-link model can provide molecular scale details.[17-21] However, this technique still requires input from experiments (such as the number of entanglements per unit volume). In principle, Molecular Dynamics (MD) only requires knowledge of intermolecular potentials. [22] Alternatively, MD can be done on a coarse-grained basis as is done in the now widely practiced Kramer-Grest approach. [23] In MD simulations flow is usually implemented by the SLLOD equations of motion. [24-26] End-bridging Monte Carlo has been applied to extensional flows, but only for unentangled chains.[27] In addition, a wide variety of successful lattice simulations capture flow effects by imposing a potential [28-33] or by probabilistically favoring displacements.[1, 34, 35]

Perhaps the most widely practiced extensional flow technique is that of the Kraynik-Reinelt boundary conditions.[36-38] When applied to a lattice, these boundary conditions allow the lattice to reproduce itself at certain discrete values of the orientation of the rotated lattice, $\theta$, and the Hencky strain is time and spatially independent. Todd and Daivis [37-40] demonstrate that this technique can be applied to real space systems. Despite their elegant mathematical nature, these simulations remain computationally intensive. In addition, the work done with these flow conditions for polymer melts[17] does not capture the zero extension limit and instead is only capable of capturing the extensional thinning regime at higher deformation rates. Because of the computational requirements, highly entangled polymer melt simulations have not been conducted.

This work draws inspiration from the work by Monaghan et. al.[41] in which the authors created a two dimensional periodic system to simulate extensional flows of Weeks-Chandler-Anderson particles. The method creates a fully periodic system that is experimentally representative of the four roll mill. [42, 43] Such a method has not been previously applied to polymer melts or three dimensional systems.

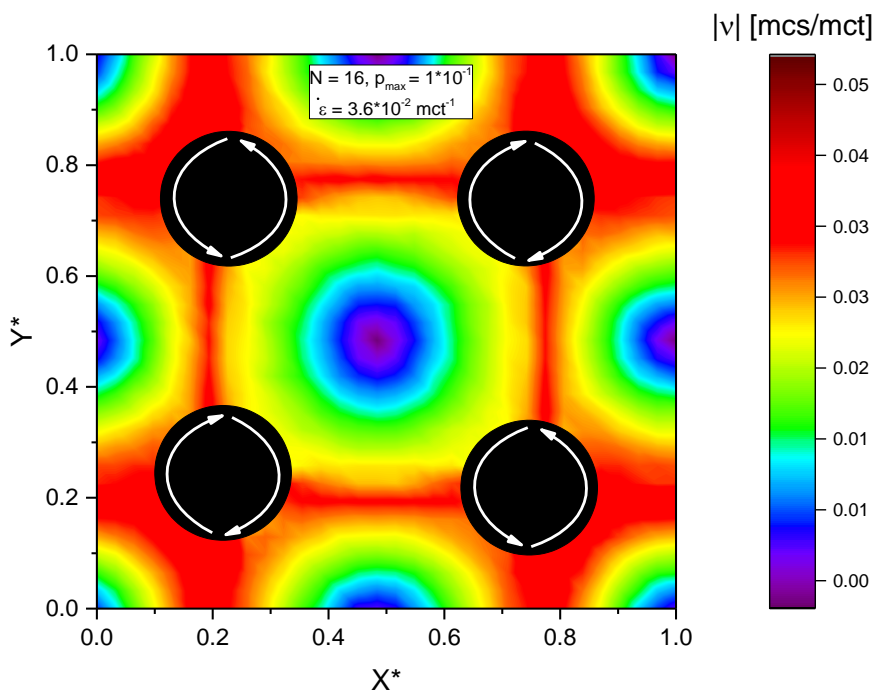

**Figure 1. (COLOR ONLINE)** Contour plot of the magnitude of velocity for an N = 16 system and a biasing parameter of $p_{max} = 1.0*10^{-1}$. Rollers are drawn in black and have arrows to indicate the direction of flow; velocity is given in Monte Carlo sites (mcs) per unit Monte Carlo time (mct).

All simulations are performed using the COMOFLO algorithm [1, 34, 35, 44-46], an adaptation of the Cooperative Motion (COMOTION) Algorithm.[47-51] This dynamic Monte Carlo simulation technique simulates polymers as coarse grained segments on a fully occupied lattice thereby allowing for realistic melt densities. This algorithm is orders of magnitude faster than other techniques but properly captures melt dynamics and rheology.[1] The algorithm has been extended to polydisperse melts,[35, 45, 46] different flow scenarios,[1, 34] hydrodynamic slip[44] and cross flow migration.[35] To capture extensional flows, the procedure described in previous works [1, 34, 35] is extended so that the attempted displacement of segments is biased in two directions. The velocity is reported in simulation units of Monte Carlo sites (mcs) per Monte Carlo time (mct); previously this conversion has been reported to be 37.7 m/s per mcs/mct.[35] Unique to the present simulations compared to previous implementations is that flow biasing in two directions is conducted.

Specifically, flow is biased to obtain hyperbolic flow representative of the four roll mill. [43] **Figure 1** provides a color contour depiction of the flow field realized. In order to simulate flow the probability of displacing a particle "forward", $p_{+y}(x,y)$ and $p_{+x}(x,y)$, or "backward", $p_{-y}(x,y)$ and $p_{-x}(x,y)$, is given by Eqns. 1-4.

$$p_{+y}(x,y) = p_{zero} + p_{max}y_{div} \quad (1)$$
$$p_{+x}(x,y) = p_{zero} - p_{max}x_{div} \quad (2)$$
$$p_{-y}(x,y) = p_{zero} - p_{max}y_{div} \quad (3)$$
$$p_{+x}(x,y) = p_{zero} + p_{max}x_{div} \quad (4)$$

Where $p_{zero}$ is the probability representative of diffusion and $p_{max}$ is the maximum biasing parameter which is proportional to the extensional deformation rate, $\dot{\varepsilon}$. When $p_{max}$ is set equal to zero displacements are equally probable in all directions and quiescent conditions are obtained. In the center of the lattice the components of velocity, $v_y(x,y)$ and $v_x(x,y)$, are given by,

$$v_y(x,y) = \dot{\varepsilon}y_{div} \quad (5)$$
$$v_x(x,y) = -\dot{\varepsilon}x_{div} \quad (6)$$

Where $\dot{\varepsilon}$ is the deformation rate, and $x_{div}$ and $y_{div}$ are a normalized coordinate that takes a value of zero in the center of simulation lattice; values of x/y=+1 correspond to the rightmost/uppermost boundary and x/y=-1 at the leftmost/bottommost boundary. **Figure 2** presents a plot of $v_x(x,y)$ and $v_y(x,y)$ as a function of the corresponding normalized box dimension for the N=16 and $p_{max}$= 0.1 case; the deformation rate is calculated by taking the derivative of the velocity with respect to position. **Figure 2** clearly shows the equality of the slopes in the two directions. This equality of the extensional deformation rate throughout the box established the validity of the approach; a purely stretching flow is obtained.

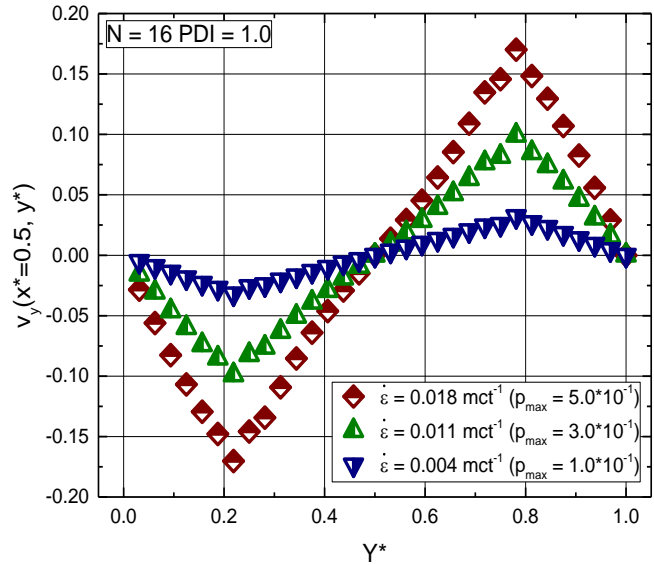

**Figure 2.** The y-component of velocity as a function of a normalized y dimension (Y*= y/total periodic image length) with the x-coordinate held constant. The velocity varies linearly with lattice position. This behavior is constant throughout the simulation lattice for both components of the velocity ensuring a purely extensional flow.

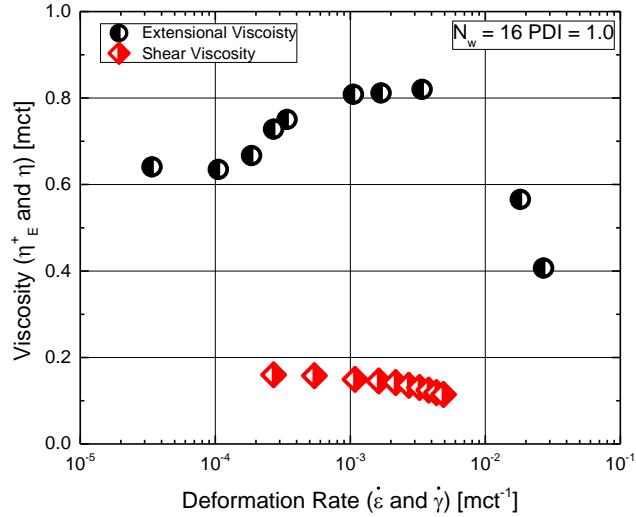

**Figure 3.** Shear and extensional viscosity of a monodisperse $N_w = 16$ system. The shear viscosity is the data previously reported by Dorgan et. al. [1] The extensional viscosity is greater than the shear viscosity and exhibits distinctly different behavior. In the low extension limit the extensional viscosity is 3.99 +/- 0.03 times the zero shear viscosity ($\eta_0 = 0.16$ +/- 0.002 mct); this compares favorably with the known limit of 4 for biaxial extension.

In order to calculate rheological properties stress must be calculated. Due to the changing direction of chain orientation in the four roll mill, the stress must be calculated using a frame of reference that moves with the chain. That is, the principle of frame invariance must be enforced.[52] To accomplish invariance, the stress is calculated along the end-to-end vector of each chain. In this frame, the z-component lies coincident with the z-direction of the periodic cell but the y-component is taken as being parallel to the end-to-end vector and the x-component is normal to the end-to-end vector. In each chain based frame of reference, the stress is calculated as the dyadic product of the bond vectors, $\langle \underline{r}\,\underline{r} \rangle$, according to,

$$\frac{\underline{\underline{\sigma}}}{2\nu kT} = \frac{2\langle \underline{r}\,\underline{r} \rangle}{3Nl^2} = \frac{2\langle \underline{r}\,\underline{r} \rangle}{3\langle R^2 \rangle} \qquad (7)$$

Where k is Boltzmann's constant, T is temperature, $v$ is the volume per segment, N is the number of segments in a chain, l is the bond length, and $\langle R^2 \rangle$ is the mean squared end-to-end vector. The extensional viscosity, $\eta_E^+$, is defined by Equation 8.

$$\eta_E^+ = \frac{\sigma_{yy} - \sigma_{xx}}{\dot{\varepsilon}} \qquad (8)$$

At low shear rates the extensional viscosity is found to be four times (4x) the zero shear viscosity as presented in **Figure 3**.

As the extension rate is increased, the viscosity departs from the low rate plateau region. There is an initial strain hardening followed by a broad maximum. The zero rate plateau is representative of the chains in the melt still being near an equilibrium coiled conformation whereas the strain hardening region is associated with the chain being extended – that is, stretched. In the final region the extensional viscosity shows a decrease with increasing extension rate (but the stress continues to increase monotonically). The simulations predict in an *a priori* manner the strain hardening followed by strain softening behavior of unentangled chains. The shape of this flow curve is consistent with experiments on unentangled polymers.[53]

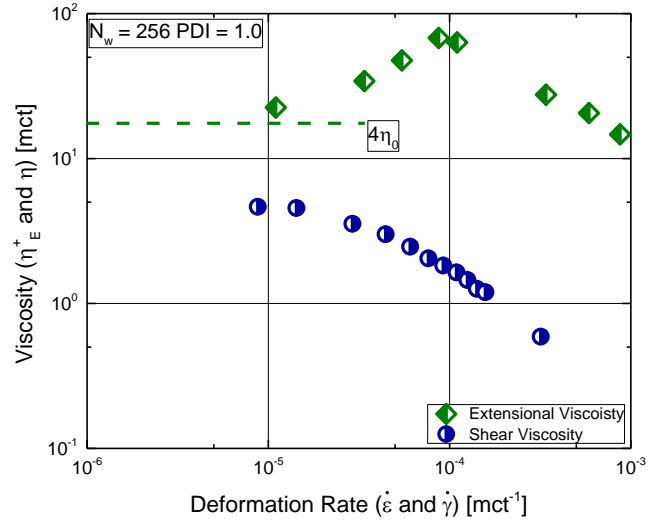

**Figure 4.** Extensional viscosity for the $N_w$ = 256, PDI = 1.0 entangled polymer melt. At low extension rates the viscosity approaches the limit for the zero extension viscosity ($\mathbf{4\eta_0}$ marked by the dotted line). At moderate extension rates there is a strain hardening phenomena leading to a maximum. The maximum is less broad than in the unentangled material of Figure 3. At high extension rates an extensional thinning region is present which scales as $\boldsymbol{\eta_E^+} \sim \dot{\boldsymbol{\varepsilon}}^{-0.693}$.

Significantly, the computational efficiency of the technique means it is capable of capturing the details of extensional flow for *entangled* polymer melts. As shown in earlier work, entanglements in the model form at a critical length corresponding to roughly $N_c$ = 100. [1] Figure 4 demonstrates that in the case of the $N_w$ = 256 melt, the Newtonian plateau is followed by an increase, maximum, and shear thinning region. It is of interest to note that this model demonstrates an extensional thinning exponent of -0.69. This value rests between that predicted by the Doi-Edwards-Marrucci-Grizzuti model (which predicts an exponent of -1) and experimental results (which demonstrate an exponent of -0.5). Recent theoretical work to include the effects of tube pressure agree with the -0.5 scaling exponent. [54] Accordingly, the issue that emerges is what is the effect of both polydispersity and average molecular weight on the shear thinning region of extensional viscosity? Clearly, the non-entangled results of Figure 2 demonstrate a much stronger dependence – an observation that is consistent with polystyrenes dissolved in its own oligomers.[55] One possibility is that the sample molecular weight must be considerably greater than the entanglement molecular weight to observe the -0.5 scaling. Also, the effects of polydispersity have yet to be clearly elucidated either experimentally or in simulations. Fortunately, the present methodology can be used to study polydisperse systems.[45]

Extension of the COMOFLO algorithm to capture extensional flow rheology has been demonstrated to provide an *a priori* and parameter free prediction of polymer melt flows. Accordingly, predictive extensional viscosities for dense entangled polymer melts are available for the first time. Not only does the algorithm capture the details of extensional flow in three dimensions, but it is able to ascertain extensional properties of entangled polymer melts across multiple flow regimes including the Newtonian plateau (where the correct ratio to the zero shear rate viscosity is exhibited), the strain-hardening regime, and the high deformation rate strain softening region. Because the technique is applicable to polydisperse systems, the ability to predict in an *a priori* manner both shear and elongational rheology for linear polymer melts of arbitrary molecular weight distribution can be considered a solved problem.

This work was funded by the Fluid Dynamics Program of the National Science Foundation under grant CBET-1067707. In addition, this material is based upon work supported by the U.S. Department of Energy, Office of Science, Office of Workforce Development for Teachers and Scientists, Office of Science Graduate Student Research (SCGSR) program. The SCGSR program is administered by the Oak Ridge Institute for Science and Education for the DOE under contract number DE-AC05-06OR23100.